\documentclass{emulateapj}

\usepackage{amssymb}
\usepackage{amsmath}
\usepackage{epsfig}
\usepackage{graphicx}
\usepackage{natbib}

\slugcomment{accepted by ApJ Letters July 2, 2010}

\begin{document}

\title{
Water/Icy Super-Earths: Giant Impacts and Maximum Water Content
}
\author{Robert A. Marcus\altaffilmark{1,a}, Dimitar Sasselov\altaffilmark{1}, 
Sarah T. Stewart\altaffilmark{2}, Lars Hernquist\altaffilmark{1}
}
\affil{$^1$Astronomy Department, Harvard University, Cambridge, MA 02138, USA}
\affil{$^2$Department of Earth and Planetary Sciences, Harvard University, 
Cambridge, MA 02138, USA}
\email{$^a$rmarcus@cfa.harvard.edu}

\begin{abstract}
  Water-rich super-Earth exoplanets are expected to be common. We
  explore the effect of late giant impacts on the final bulk abundance
  of water in such planets. We present the results from smoothed
  particle hydrodynamics simulations of impacts between differentiated
  water(ice)-rock planets with masses between 0.5 and 5 M$_{\oplus}$
  and projectile to target mass ratios from 1:1 to 1:4. We find that
  giant impacts between bodies of similar composition never decrease
  the bulk density of the target planet. If the commonly assumed
  maximum water fraction of 75wt\% for bodies forming beyond the snow
  line is correct, giant impacts between similar composition bodies
  cannot serve as a mechanism for increasing the water fraction.
  Target planets either accrete materials in the same proportion,
  leaving the water fraction unchanged, or lose material from the
  water mantle, decreasing the water fraction. The criteria for
  catastrophic disruption of water-rock planets are similar to those
  found in previous work on super-Earths of terrestrial
  composition. Changes in bulk composition for giant impacts onto
  differentiated bodies of any composition (water-rock or rock-iron)
  are described by the same equations. These general laws can be
  incorporated into future N-body calculations of planet formation to
  track changes in composition from giant impacts.

\end{abstract}


\section{Introduction}
Super-Earths, massive terrestrial exoplanets in the regime $\lesssim$
10 M$_{\oplus}$, are expected to come in a diversity of bulk
compositions. Both planet formation theory and comparative planetology
stand to benefit from distinguishing super-Earths of different
composition. It is especially valuable to know how water-rich such
planets could be. Water is abundant in protoplanetary disks, but it is
also very volatile, so the details of the planet formation (and
post-formation) process can be crucial in determining the distribution
of water-rich planets by bulk composition and orbit.

The catalog of observed extrasolar planets now includes more than 400
members.\footnote{see http://exoplanet.eu} Of these, more than 70
planets have been observed transiting their parent star. Until
recently, the only known transiting planets were similar in mass to
either Jupiter or Neptune.  With CoRoT-7b \citep{Leger:2009,
  Queloz:2009} and GJ1214b \citep{Charbonneau:2009}, we now have the
first observed transits of planets in the super-Earth mass regime.
Many more detections are expected in the near future from the {\it Kepler}
mission \citep{Kepler:2009}. A transit observation provides a
determination of the planet's radius as well as the inclination angle
of the planet's orbit with respect to the line of sight. When combined
with the mass determined from radial velocity measurements, the mean
density of the planet is derived.

Given the bulk density of a super-Earth, it is possible to estimate a
range of possible internal compositions, from rocky/iron terrestrial
planets with small radii for a given mass, to icy and gas rich planets
with much larger radii \citep[e.g.,][]{Valencia:2006, Valencia:2007b,
  Fortney:2007, Seager:2007}. The precise composition of the planet
cannot be determined from internal structure models alone because of
the degeneracy in the mass--radius relationship for various
compositions.  However, with precise measurements of radius and mass
(uncertainties $\lesssim$ 5\% and 10\%, respectively), one can
distinguish an icy/rocky super-Earth from a rocky/iron super-Earth
\citep{Valencia:2007b}.

The mass and radius measurements for CoRoT-7b, 4.8 $\pm$
0.8M$_{\oplus}$ and 1.68 $\pm$ 0.09$R_{\oplus}$, respectively, are
consistent with a planet of terrestrial composition \citep{Leger:2009,
  Queloz:2009}. GJ1214b, with a similar mass of 6.55 $\pm$
0.98$M_{\oplus}$, has a radius of 2.678 $\pm$ 0.13$R_{\oplus}$,
implying the presence of a thin hydrogen/helium envelope around a core
composed primarily of ice or, alternatively, a thick hydrogen/helium
envelope around a rocky core \citep{Charbonneau:2009}.

Further constraints on the compositions of transiting super-Earths can
be obtained from a theoretical understanding of planet formation.
\cite{Ida_LinI} and \cite{Mordasini:2009a} have applied detailed
models of planet formation to the generation of synthetic populations.
\cite{Figueira:2009} have combined the planet formation model of
\cite{Mordasini:2009a} with simplified interior structure models to
determine a statistical distribution of possible compositions for the
transiting hot Neptune GJ436b, excluding the possibilities that the
planet is either an envelope-free water planet or a rocky planet with
a hydrogen/helium envelope. Such constraints represent a major triumph
for planet formation models. However, all current models neglect the
potentially important effects of multiple embryo formation and giant
impacts.

Giant impacts are thought to have played a major role in shaping the
planets in our solar system. Such events have been shown to explain
the high bulk density of Mercury in the inner solar system
\citep{Benz:1988, Benz:2007}, the formation of the Haumea collisional
family in the outer solar system \citep{Brown:2007,Leinhardt:2010},
and properties of several planets in between. In fact, giant impacts
are an inevitable consequence of the core accretion model of planet
formation \citep[e.g.,][]{Wetherill:1994}. \cite{Marcus:2009}
developed scaling relations to determine the outcome of giant impacts
between planets of terrestrial composition. They extended the
catastrophic disruption criteria of \cite{Stewart_Leinhardt:2009} into
the super-Earth mass regime and provided a power-law relationship for
the iron mass fraction of the largest impact remnant given target
mass, impactor mass, and impact velocity. These laws were then
combined with simple dynamical arguments to constrain the minimum
radius of super-Earths as a function of mass \citep{Marcus:2010a}.

Ultimately, we want to know what the distribution of super-Earths of
different bulk composition looks like in the mass-radius diagram. The
work by \cite{Marcus:2010a} defines the theoretically anticipated
lower envelope of that distribution. This Letter attempts to similarly
define an upper envelope, if one does occur, that might separate water
planets with no hydrogen/helium envelopes from mini-Neptunes with
hydrogen/helium envelopes around an icy/rocky or just rocky core. We
follow the same premise, that well-defined initial bulk compositions
can be modified by late-time giant impacts to reach either of the
envelopes of the mass-radius distribution of super-Earths.

\section{Method}

We performed a series of more than 100 simulations using the smoothed
particle hydrodynamics (SPH) code GADGET \citep{Springel:2005},
modified to read tabulated equations of state
\citep[EOS,][]{Marcus:2009}. For more details on using this scheme for
giant planetary impact simulations see \cite{Marcus:2009,
  Marcus:2010a}.
 
We considered targets ranging in mass from 0.5 $M_\oplus$ to 5
$M_\oplus$. The mass of the planets was 50\% H$_{2}$O and 50\%
serpentine (Mg$_3$Si$_2$O$_5$(OH)$_4$, $\sim$ 11-15\% H$_2$O by mass)
with a bulk density of $2.2-3.6$~g~cm$^{-3}$.  The higher bulk
densities for the largest mass planets reflect the formation of dense
high-pressure polymorphs of ice and the corresponding smaller total
planetary radii. This composition is consistent with the protosolar
ice/rock ratio of $\sim$ 2.5 \citep{Guillot:1999} and more ice-rich
than the composition of the dwarf planets and large KBOs in the outer
solar system \citep{McKinnon:2008}. The serpentine EOS was tabulated
from the code ANEOS \citep{Thompson_Lauson:1972}, with input
parameters taken from \cite{Brookshaw:1998}. The ice EOS was the
tabulated 5-Phase equation of state for H$_2$O
\citep{Senft_Stewart:2008}.  Bodies were initialized with fixed
temperature profiles from \cite{Fu:2010} and all materials are
hydrodynamic.  Each body had 25000-100000 SPH particles. A subset of
the simulations was run at twice this resolution, with identical
results.

The projectile-to-target mass ratios were 1:4, 1:2, and 1:1. Because
the outcome of giant impacts has been shown to depend very strongly on
impact angle \citep{Agnor_Asphaug:2004, Marcus:2009}, we considered
impact parameters of $b$=0 (0$^{\circ}$, head-on), $b$=0.5
(30$^{\circ}$), and $b$=0.707 (45$^{\circ}$).

\section{Results}

\subsection{Catastrophic Disruption Criteria and Icy Mantle Stripping}

\begin{figure}
\figurenum{1}
\begin{center}
\includegraphics[scale=0.6]{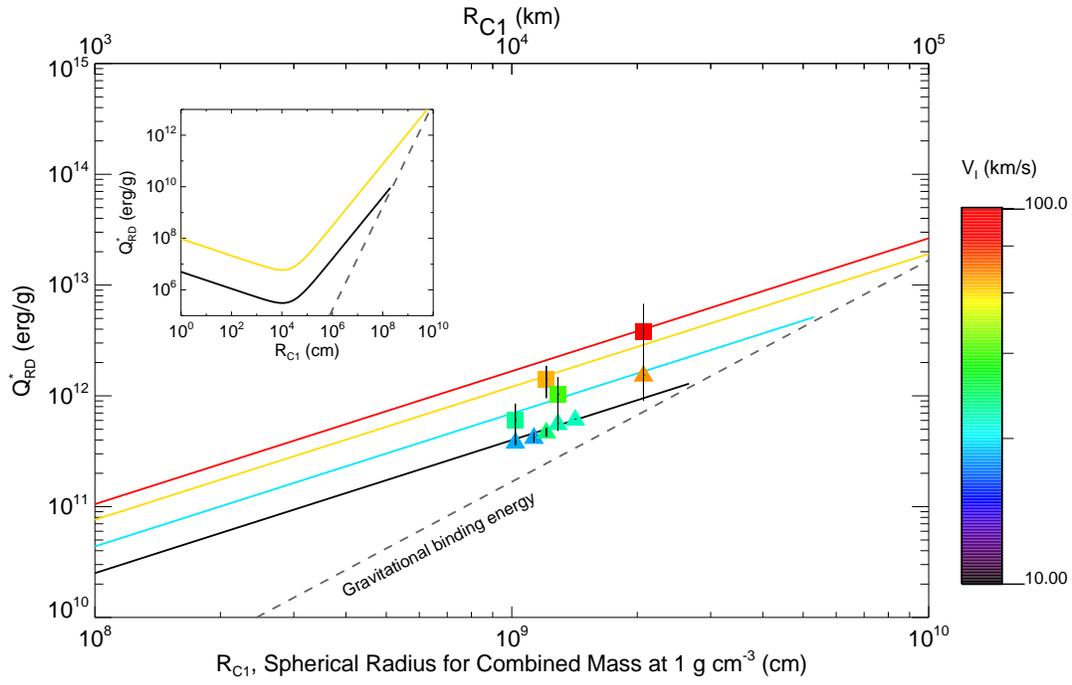}
\end{center}
\caption{Catastrophic disruption criteria $Q^{*}_{\rm{RD}}$
  vs. $R_{\rm{C1}}$ (Equation (\ref{eq:qsrd}) for $V_{\rm{i}}=$10, 20,
  40, 60 km~s$^{-1}$) and hydrocode results (triangles: impact
  parameter $b$ = 0; squares: $b$ = 0.5).  Inset shows both strength-
  and gravity-dominated regimes for the size and velocity-dependent
  $Q^{*}_{\rm{RD}}$.}
\label{fig:Q*RD}
\end{figure}

Giant collisions that are near or directly head-on (zero impact
parameter $b$) of sufficient energy lead to disruption and net
erosion. The criterion for catastrophic disruption is an impact
condition that leads to a largest remnant with half the original
combined colliding mass.  In the disruption regime, the mass of the
largest impact remnant is a function of the reduced mass kinetic
energy divided by the total colliding mass, scaled by the catastrophic
disruption criteria $Q^{*}_{\rm{RD}}$, a scaling factor that depends
primarily on the total colliding mass and the impact velocity. The
reduced mass kinetic energy, scaled to the total colliding mass is
$Q_{\rm{R}} = \frac{1}{2}\mu V_{\rm{i}}^{2}/M_{\rm{tot}}$, where $\mu$
is the reduced mass, $V_{\rm{i}}$ is the impact velocity, and
M$_{\rm{tot}}$ is the total mass. The size- and velocity-dependent
catastrophic disruption criteria for weak bodies in the
gravity-dominated regime are given by \citep[][Equation
  2]{Stewart_Leinhardt:2009}
\begin{equation}
  Q^{*}_{\rm{RD}} = 10^{-4} R_{\rm{C1}}^{1.2} V_{\rm{i}}^{0.8}.
  \label{eq:qsrd}
\end{equation}
Here $R_{\rm{C1}}$ is the radius of a spherical body with mass
$M_{\rm{tot}}$ and a mean density of 1 g cm$^{-3}$. These disruption
criteria, originally fit to $N$-body impact simulations between
planetesimals, were shown by \cite{Marcus:2009} to extend into the
super-Earth mass regime for bodies of terrestrial composition. In
Figure \ref{fig:Q*RD}, the fitted $Q^{*}_{\rm{RD}}$ values for
collisions between icy planets are shown as a function of
$R_{\rm{C1}}$. The disruption criteria of
\cite{Stewart_Leinhardt:2009} fit the data for ice-rock planets as
well, predicting $Q^{*}_{\rm{RD}}$ to within a factor of 2 of the
hydrocode calculations for head-on ($b$=0) impacts. The predicted
$Q^{*}_{\rm{RD}}$ is smaller than the hydrocode calculations for
head-on impacts; however, the hydrocode calculations do not include
material strength which biases the value of $Q^{*}_{\rm{RD}}$ downward
\citep[e.g., see][]{Leinhardt_Stewart:2009}.  Oblique impacts are
discussed below. Thus, these disruption criteria are largely
independent of the internal compositions of the colliding bodies.

\begin{figure}
\figurenum{2}
\begin{center}
\includegraphics[scale=0.6]{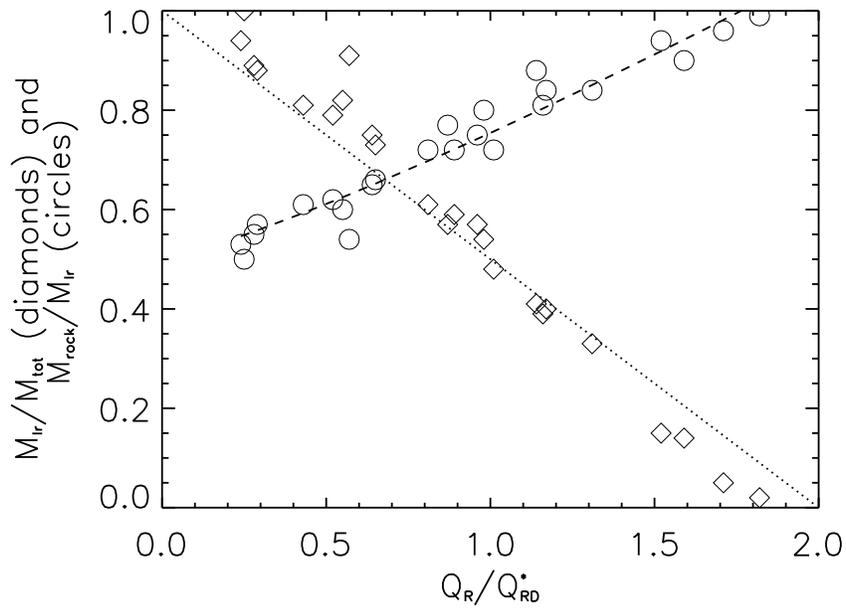}
\end{center}
\caption{Largest remnant mass and core mass fraction vs. scaled impact
  energy for head on impacts. The dotted line is Equation
  (\ref{eq:mlreq}). The dashed line is Equation
  (\ref{eq:core_fraction}).}
\label{fig:disruption}
\end{figure}

\cite{Stewart_Leinhardt:2009} found a linear relationship between
$Q_{\rm{R}}$ and the mass of the largest remnant scaled to the total
colliding mass:
\begin{equation}
  M_{\rm{lr}}/M_{\rm{tot}} = -0.5(Q_{\rm{R}}/Q^{*}_{\rm{RD}} -1) + 0.5 \, .
  \label{eq:mlreq}
\end{equation}
In Figure \ref{fig:disruption}, the mass of the largest remnant,
scaled to the total colliding mass, is shown as a function of the
reduced mass kinetic energy. The linear relationship still holds for
super-Earth planets consisting largely of ice (dotted line). Thus,
this scaling relationship should be valid for all solid planetary
bodies, regardless of the details of the internal structure.

In these disruptive simulations, some of the icy mantle is removed,
leading to an increase in bulk density. The rocky core mass fraction
of the largest impact remnant is given by a power law of
$Q_{\rm{R}}$/$Q^{*}_{\rm{RD}}$,
\begin{equation} 
  M_{\rm{core}}/M_{\rm{lr}} = 0.5 + 0.25(Q_{\rm{R}}/Q^{*}_{\rm{RD}})^{1.2},
  \label{eq:core_fraction}
\end{equation} 
shown by the dashed line in Figure \ref{fig:disruption}.

\subsection{Effect of the impact parameter: Disruption versus Hit and Run}

High velocity head-on impacts fall in the disruption regime discussed
above. As the impact parameter increases and the impacts become more
grazing, a different regime emerges. The hit-and-run impact events
described by \cite{Agnor_Asphaug:2004} and \cite{Marcus:2009} also
occur for the planets presently considered. This regime is defined by
a sharp discontinuity between merging impacts at low velocity and high
velocity collisions in which the impactor grazes the target and both
bodies escape the encounter largely intact.

\begin{figure}
\figurenum{3}
\begin{center}
\includegraphics[scale=0.6]{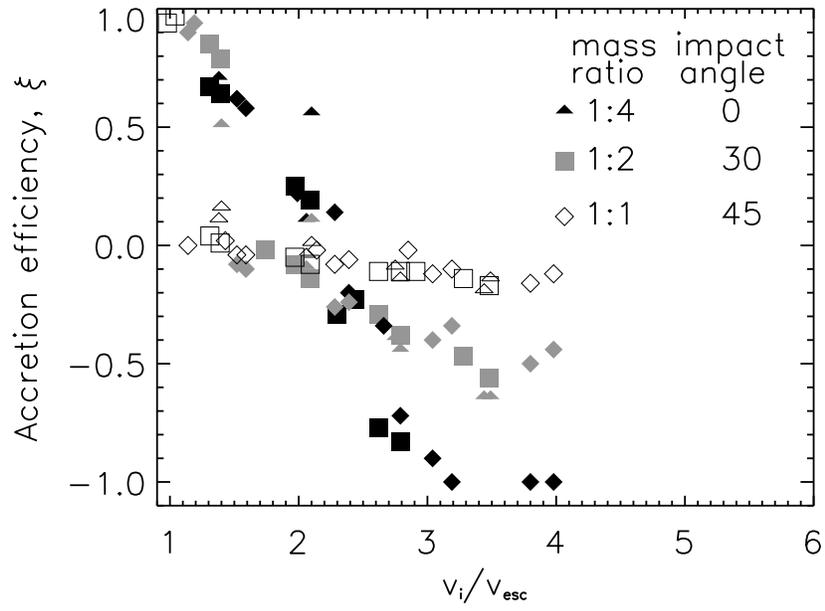}
\end{center}
\caption{Accretion efficiency $\xi$ (Equation (\ref{eq:xi}))
  vs. scaled impact velocity. The symbol denotes the mass ratio and
  the shading denotes the impact angle.}
\label{fig:accretion}
\end{figure}

Figure \ref{fig:accretion} shows the accretion efficiency, $\xi$, as a
function of the impact velocity scaled to the mutual escape velocity
($v_{\rm{esc}}^{2} =
2G(M_{\rm{targ}}+M_{\rm{proj}})/(R_{\rm{targ}}+R_{\rm{proj}})$). The
accretion efficiency is defined as the difference between the mass of
the largest impact remnant and the target mass, divided by the mass of
the impactor \citep{Asphaug:2009}:
\begin{equation}
\xi \equiv \frac{M_{\rm{lr}} - M_{\rm{targ}}}{M_{\rm{proj}}}.
\label{eq:xi}
\end{equation}
The impact velocity of the transition between accretion and
hit-and-run depends on the impact angle and the projectile-to-target
mass ratio, but always falls in the range
$v_{\rm{i}}\sim1.1-1.5v_{\rm{esc}}$. For the icy planets, all of the
$b$=0.7 simulations transitioned from merging to hit-and-run and
remain in the hit and run regime for the range of impact velocities
considered here. The simulations with $b$=0 smoothly transitioned from
merging to disruption. For $b$=0.5, there is a small hit-and-run
regime between merging and disruption.

In the hit-and-run regime, catastrophic disruption may still occur at
extremely high impact velocities. Figure \ref{fig:Q*RD} includes the
calculated catastrophic disruption criteria versus combined spherical
radius for 30$^{\circ}$ impacts with mass ratios of 1:2 and 1:4; the
predicted $Q^{*}_{\rm{RD}}$ using Equation (\ref{eq:qsrd}) are only
20\% larger on average when accounting for the different critical
impact velocities. The 30$^{\circ}$ impacts with mass ratio of 1:1 and
all the 45$^{\circ}$ impacts resulted in hit and run up to
$4V_{\rm{esc}}$, and we could not derive a $Q^{*}_{\rm{RD}}$.

30$^{\circ}$ impact conditions typically result in $Q^{*}_{\rm{RD}}$
values larger than that for head-on impacts by a factor of $\sim$2.
However, there is also a non-trivial dependence on the mass-ratio of
the colliding bodies. The combined factors of mass ratio and impact
angle are complicated enough to warrant a future dedicated study for
the boundary between the disruption and hit-and-run regimes.

\subsection{Two Models for Mantle Stripping in a Differentiated Body}

Combining the results of the simulations of collisions between
rock/ice super-Earths discussed above with the collision simulations
of iron/rock super-Earths from \cite{Marcus:2009}, we construct two
general models for collisional mantle stripping in a two material,
differentiated planet. In the first one, we begin by calculating
$Q^{*}_{\rm{RD}}$ from Equation (\ref{eq:qsrd}), followed by the
expected mass of the largest remnant from Equation
(\ref{eq:mlreq}). In this model, we assume that all escaping mass is
the lightest material; that is, the cores of the two bodies merge. If
the combined mass of the two cores is less than the mass of the
largest remnant, mantle material is added until the total mass reaches
$M_{\rm{lr}}$. Given the core masses of the target and projectile,
$M_{\rm{core,targ}}$ and $M_{\rm{core,proj}}$, the mass of the final
post-impact core is then
\begin{equation}
  M_{\rm{core}} = \rm{min}(M_{\rm{lr}}, M_{\rm{core,targ}}+M_{\rm{core,proj}}).
  \label{eq:model1}
\end{equation}

In the second model, the first step is again to calculate
$Q^{*}_{\rm{RD}}$ and the expected mass of the largest remnant from
Equations (\ref{eq:qsrd}) and (\ref{eq:mlreq}), respectively. In the
accretion regime, where the mass of the largest remnant is larger than
the target mass, we assume that core material is accreted first from
the impactor followed by mantle material. The mass of the final
post-impact core is then
\begin{equation}
  M_{\rm{core}} = M_{\rm{core,targ}} + \rm{min}(M_{\rm{core,proj}},
  M_{\rm{lr}} - M_{\rm{targ}}).
  \label{eq:model2acc}
\end{equation}
In the disruption regime, where the mass of the largest remnant is
smaller than the target mass, we assume that the material is stripped
away from the target in spherical shells, beginning with mantle
material only. The core mass of the largest remnant is
\begin{equation}
  M_{\rm{core}} = \rm{min}(M_{\rm{core,targ}}, M_{\rm{lr}}).
  \label{eq:model2dis}
\end{equation}

\begin{figure}
\figurenum{4}
\begin{center}
\includegraphics[scale=0.6]{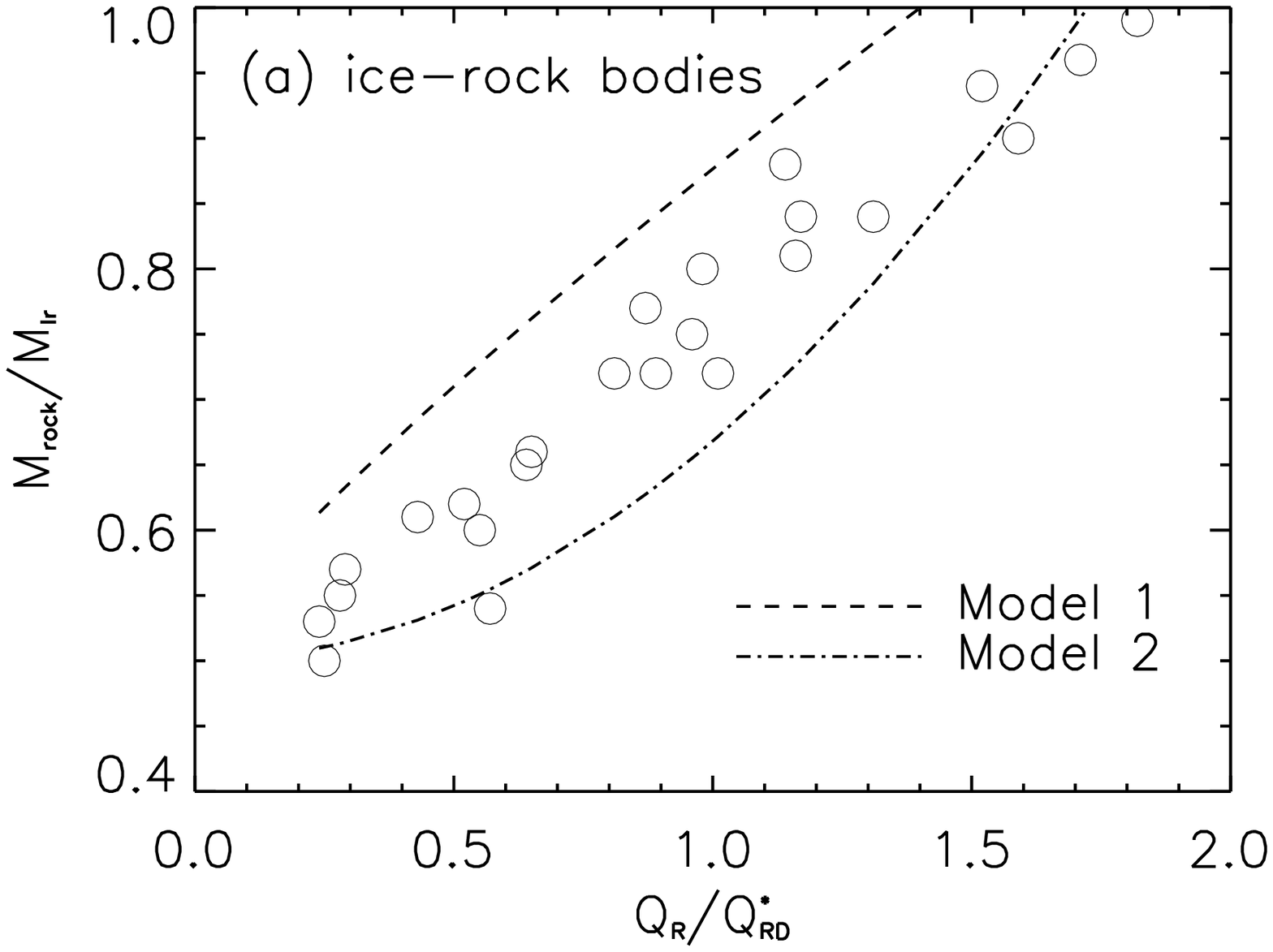}
\includegraphics[scale=0.6]{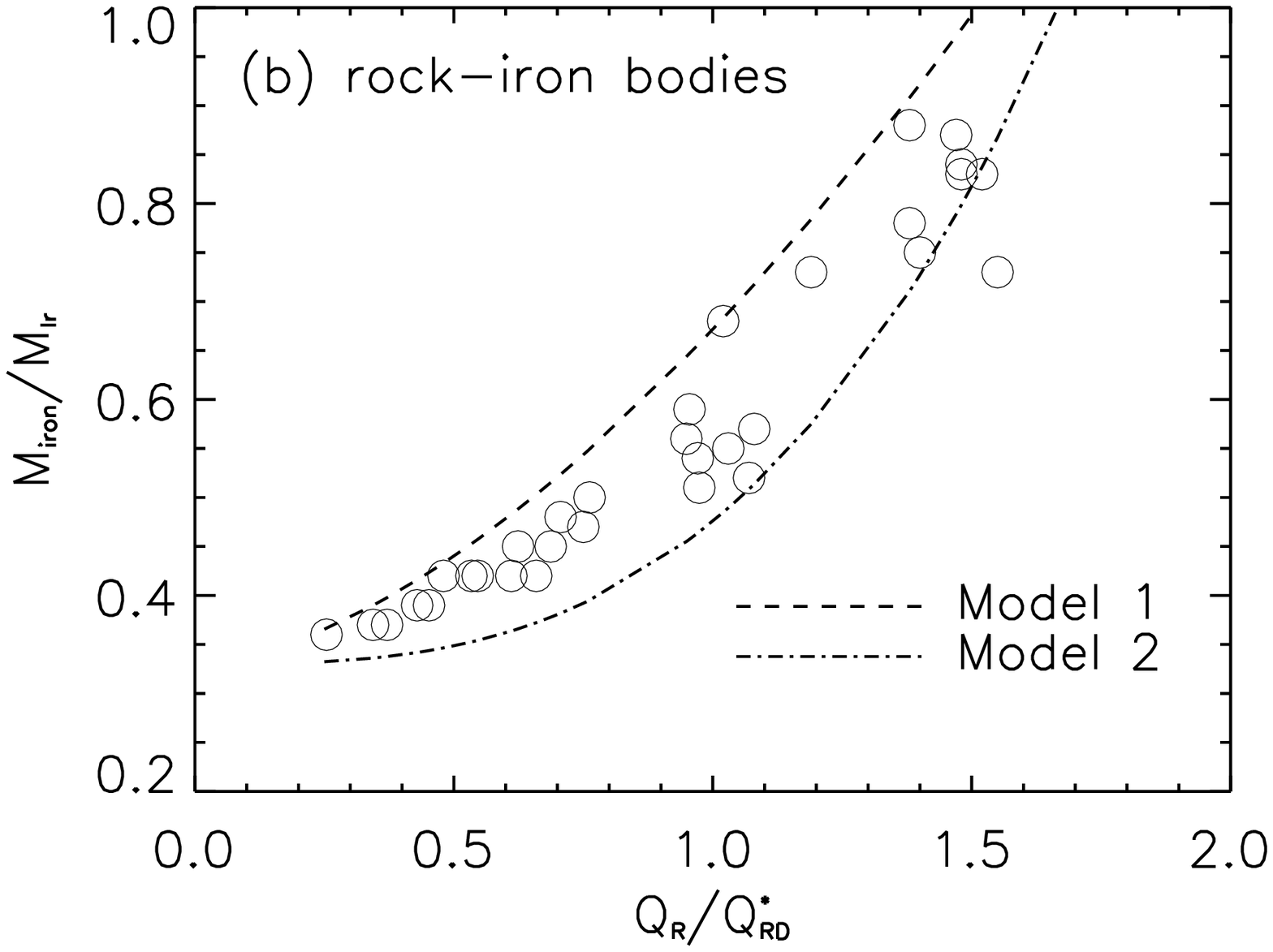}
\end{center}
\caption{Core mass fraction of the largest remnant vs. scaled impact
  energy for head on impacts. Panel (a) is for ice-rock bodies and
  panel (b) is for rock-iron bodies \citep{Marcus:2009}. The dashed
  line is the prediction of model 1 (Equation (\ref{eq:model1})). The
  dash-dotted curve is the prediction of model 2 (Equations
  (\ref{eq:model2acc}) and (\ref{eq:model2dis})). The circles are
  simulation results.}
\label{fig:model}
\end{figure}

A comparison in core mass fraction between these two simple models of
accretion/stripping and the simulation results is shown in Figure
\ref{fig:model}, for both ice-rock bodies (a) and rock-iron bodies (b)
\citep{Marcus:2009}. The dashed curve is the prediction from the first
model, the dash-dotted curve is the prediction from the second model,
and the points are simulation results. The first model generally
overpredicts the core mass fraction, while the second model generally
results in an underprediction. Given the assumptions of the two
models, this result is unsurprising. The first model assumes that both
cores are incorporated into the final body, even in the disruption
regime. The second model ignores any loss of material from the target
mantle in the accretion regime and that assumes material is lost in
shells in the disruption regime. In reality, some material will be
lost from the target mantle even in the accretion regime, and in the
disruption regime, material is preferentially lost from the side of
the target suffering the impact, including the core when all mantle
material is stripped.

Despite these discrepancies, the models match the simulations fairly
well, to within $\sim$15\% for ice-rock bodies and $\sim$10\% for
rock-iron bodies. Note that the calculations for ice-rock planets were
checked for sensitivity to the number of particles in the simulation.

These simple models allow for conservative estimates of the limiting
case of maximizing the total water content of the largest remnant
after a giant impact. While Figure \ref{fig:model} shows the results
for head-on impacts only, the model holds also for off-axis impacts in
the merging and disruption regimes. Hit-and-run impacts result in very
little material exchange and only minor collisional erosion of the
target and projectile mantles.

\section{Discussion and Conclusions}

In previous work, we demonstrated that the catastrophic disruption
criteria of \cite{Stewart_Leinhardt:2009} extend into the super-Earth
mass regime for rock-iron differentiated bodies. Here, we show that
the disruption criteria still hold for bodies of ice-rock
composition. Thus, the catastrophic disruption criteria are largely
independent of internal composition of the colliding bodies, holding
for gravitationally bound rubble piles as well as for icy
super-Earths. The catastrophic disruption criteria (Equation
(\ref{eq:qsrd})) and universal law for the mass of the largest remnant
(Equation (\ref{eq:mlreq})) are a powerful tool for determining
collisional outcomes that can easily be incorporated into $N$-body
planet formation models.

In the disruption regime, we derive a scaling law for the mass
fraction of the rocky core following collisions between these icy
planets. While this relation, as well as the similar law for iron mass
fraction following collisions between rock-iron planets derived in
\cite{Marcus:2009}, is interesting, we have extended the usefulness of
such relationships by showing that a more general law holds for how
the composition changes in giant impacts between differentiated
bodies. In our simple model, collisions that lead to accretion
initially incorporate the heavier materials from the impactor while
erosive collisions remove the lightest material from the target
first. This simple picture agrees to within $\sim$10\% with the
simulations for rock-iron bodies and is accurate to $\sim$15\% for
ice-rock bodies. This result will be very useful to $N$-body planet
formation models that track the composition of growing planets
\citep[e.g.,][]{Raymond:2006, Thommes:2008}

Further, our current results, as well as those in \cite{Marcus:2009},
show that giant impacts between bodies of similar composition will
never serve to decrease the bulk density of the target planet. If the
commonly assumed maximum ice fraction of 75 wt\% for bodies forming
beyond the snow line \citep[e.g.,][]{Mordasini:2009a} is correct,
giant impacts cannot serve as a mechanism for increasing the ice
fraction.  Such impacts would either accrete heavier materials in the
same proportion as in the target, leaving the ice fraction unchanged,
or would strip away material from the target's icy mantle, decreasing
the ice fraction. This result confirms scenario (3) for the initial
states of super-Earth planets by \citet[][impacts can either strip the
mantle or provide late water delivery in the proportions of the
projectile]{Valencia:2007b} and a maximum radius for water
super-Earths in the mass--radius diagram.

One scenario that would seemingly violate this rule and result in very
underdense planets would be a collision similar to the Moon-forming
impact \citep{Canup:2004}, followed by a dynamical event that led to
the satellite becoming a planet, unbound from its parent body. This
planet would then be highly depleted in core material. However, such
satellites forming from impact-created disks are typically not more
than a few percent of the mass of the parent body \citep{Canup:2001}
and would thus not fall in the super-Earth mass regime. Massive
secondary bodies survive only in hit-and-run events, in which case
there is only minor collisional erosion of the target and projectile
mantles.

Therefore, when considering the possible internal compositions for
transiting super-Earths with relatively large radii, such as GJ1214b
\citep{Charbonneau:2009}, a composition of about 75 wt\% ice is the
highest ice fraction that is consistent with our current understanding
of planet formation. If the planet's radius is larger than this, the
presence of a gaseous hydrogen--helium envelope should be assumed. In a
sense, the curve on the mass--radius relationship corresponding to
75 wt\% ice is then a likely upper-bound on the radii for terrestrial
planets.  Above this curve, planets should be similar to the ice
giants of the solar system.

Not surprisingly, the impact conditions necessary to remove a
substantial fraction of an icy mantle are much less extreme than those
necessary to remove silicate mantles \citep{Marcus:2010a}. While
removing an entire rocky mantle in a giant impact is unlikely because
of the large impact velocities necessary, the impact velocities needed
to strip much of an icy mantle are attainable during planet
formation. Applying the results of this Letter and previous work
\citep{Marcus:2009, Marcus:2010a}, future planet formation models
incorporating the scaling laws for giant impacts can accurately track
compositional changes during planet formation. Work analogous to that
of \cite{Mordasini:2009a} can then provide us with statistical
likelihoods for such icy mantle-removing impacts during late stage
giant impacts.

The NASA {\it Kepler} mission should uncover an unprecedented large number
of transiting super-Earth planets \citep{Borucki:2010}. We expect that
many of them will have masses $M_p \leq 5 M_E$ and relatively short
orbits, ensuring a negligible fraction of planets with H/He
envelopes. If this expectation turns out to be true, the {\it Kepler} data
should show a clear upper envelope in the mass--radius
distribution. Otherwise, atmospheric spectroscopy will be needed to
distinguish the ``mini-Neptunes'' \citep{Miller-Ricci:2009}.

The simulations in this Letter were run on the Odyssey cluster
supported by the Harvard FAS Research Computing Group.

\clearpage

\clearpage

\clearpage

\clearpage 

\clearpage

\bibliographystyle{apj_fixed}

\end{document}